\documentclass[preprint,preprintnumbers, prd, floatfix, superscriptaddress,nofootinbib,amsmath]{revtex4-1} %,tightenlines
\usepackage{graphicx}
\usepackage{epsfig}
\usepackage{amsmath}
\usepackage{subfigure}
\usepackage{dcolumn}
\usepackage{bm}
\usepackage[usenames ,dvipsnames]{xcolor}
\usepackage{slashed}
\usepackage{graphicx,color}
\definecolor{darkerblue}{rgb}{0, 0, 1}
\usepackage[bookmarksnumbered, pdfstartview=FitH,colorlinks,urlcolor=darkerblue, citecolor=darkerblue,linkcolor=darkerblue] {hyperref}
\usepackage{appendix}

\begin{document}
\title{A phenomenological estimate of rescattering effects in $B_s\to K^{*0}\bar{K}^{*0}$}
\author{Yao Yu}
\email{Corresponding author: yuyao@cqupt.edu.cn}
\affiliation{Chongqing University of Posts \& Telecommunications, Chongqing, 400065, China}
\affiliation{Department of Physics and Chongqing Key Laboratory for Strongly Coupled Physics, Chongqing University, Chongqing 401331, People's Republic of China}
\author{Hai-Bing Fu}
\email{Corresponding author: fuhb@gzmu.edu.cn}
\affiliation{Department of Physics, Guizhou Minzu University, Guiyang 550025, P.R.China}
\author{Han Zhang}
\email{Corresponding author: zhanghanzzu@gs.zzu.edu.cn}
\affiliation{School of Physics and Microelectronics,
Zhengzhou University, Zhengzhou, Henan 450001, China}
\author{Bai-Cian Ke}
\email{Corresponding author: baiciank@ihep.ac.cn}
\affiliation{School of Physics and Microelectronics,
Zhengzhou University, Zhengzhou, Henan 450001, China}
%------------------------------------------------------------------------------
\begin{abstract}
  The measurements in $b\to s$ penguin-dominated decays are widely recognized
  as a powerful test for searching for New Physics by studying the deviation
  from theoretical estimations within the Standard Model. We examine the
  final-state rescattering effects on the decay $B_s\to K^{*0}\bar{K}^{*0}$
  and provide estimations of the branching ratio and longitudinal polarization
  of $B_s\to K^{*0}\bar{K}^{*0}$, which is consistent with experimental
  observations. Our conclusion is that both short- and long-distance
  interactions contribute significantly in this decay. The small
  longitudinal polarization in $B\to VV$ modes may not be a signal for New
  Physics.
\end{abstract}
\maketitle
%------------------------------------------------------------------------------
\section{Introduction}
The measurement in $b\to s$ processes have been widely recognized as a nice way
to investigate quantum chromodynamics~(QCD), $CP$ violation, and potential new
physics~(NP) beyond the Standard
Model~(SM)~\cite{Grossman:1996ke,London:1997zk,Grossman:2024amc,Ciuchini:2012gd,Bhattacharya:2012hh,Ciuchini:2007hx,Gronau:1994rj,Grossman:1997gr,Amhis:2022hpm}. Among these,
$B_s\to K^{*0}\bar{K}^{*0}$ is regarded as a good mode sensitive to
NP~\cite{Grossman:1997gr,Ciuchini:2007hx}. The latest experimental average for
the branching ratio and longitudinal polarization are
${\cal B}(B_s\to K^{*0}\bar{K}^{*0})=(11.1\pm 2.7)\times10^{-6}$ and
$f_L(B_s\to K^{*0}\bar{K}^{*0})=0.240\pm0.031\pm0.025$~\cite{ParticleDataGroup:2022pth}.
Various theoretical approaches provide predictions within the SM for
${\cal B}(B_s\to K^{*0}\bar{K}^{*0})$ and $f_L(B_s\to K^{*0}\bar{K}^{*0})$,
such as QCD factorization~(QCDF)~\cite{Beneke:2006hg,Chang:2017brr},
Perturbative QCD (PQCD)~\cite{Zou:2015iwa,Yan:2018fif}, Soft-Collinear
Effective Theory~(SCET)~\cite{Wang:2017rmh}. Nevertheless, the estimations are
commonly smaller than the experimental observations for the branching ratio and
larger for longitudinal polarization.
%------------------------------------------------------------------------------

\begin{figure}[t!]
\includegraphics[width=0.40\textwidth]{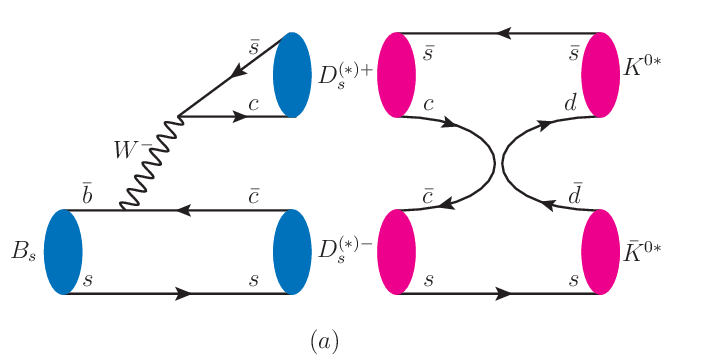}
\includegraphics[width=0.35\textwidth]{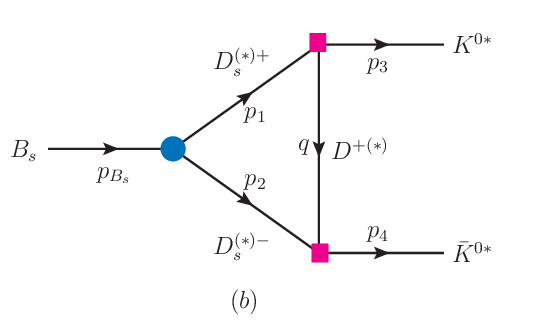}
\caption{Rescattering $B_s\to K^{*0}\bar{K}^{*0}$ decays involve eight
  Feynman diagrams: (a) depicts the process at the quark level, while (b)
  illustrates it at the hadron level.}\label{ku}
\end{figure}

To address these discrepancies between theoretical predictions and experimental
measurements, several studies have focused on investigating potential
contribution from
NP~\cite{Geng:2021lrc,Li:2022mtc, Lizana:2023kei, Descotes-Genon:2011rgs, Bhattacharya:2012hh}.
These new physics models typically add a non-universal $Z^\prime$ boson, scalar
leptoquarks, or other new types of particles to the SM, thereby fitting
theoretical results to match experimental data. The general characteristic of
the aforementioned theories is based on the factorization hypothesis, focusing
mainly on short-distance interactions and ignoring the contributions of the
non-factorizable parts, i.e., long-distance interactions. However, the
contribution from NP may be excluded by that from long-distance interactions
within the SM. At quark level, $\bar{b}\to d\bar{d}\bar{s}$ can receive
contribution through $\bar{b}\to c\bar{c}\bar{s}\to d\bar{d}\bar{s}$, where
$c\bar{c}\bar{s}\to d\bar{d}\bar{s}$ proceeds via the strong interaction as
shown in Fig.~\ref{ku}(a). At the hadron level,
$B_s\to D_s^{(*)+}D_s^{(*)-}$ decays are followed by the $D_s^{(*)+}D_s^{(*)-}$
to $K^{*0}\bar{K}^{*0}$ rescattering via exchange of a $D^{(*)+}$ meson
depicted in Fig.~\ref{ku}(b). The effect of this ``triangle-rescattering''
heavily depends on the couplings of the intermediate interactions involved.
The branching ratios of $B_s\to D_s^{(*)+}D_s^{(*)-}$ have been measured at
$10^{-2}$ level, and the strong couplings of
$D_s^{(*)+}\to D^{(*)+}K^{*0}$~($D_s^{(*)-}\to D^{(*)-}\bar{K}^{*0}$) have
been identified as substantial~\cite{Cheng:2004ru, Wu:2023fyh}. Hence, the
$B_s\to K^{*0}\bar{K}^{*0}$ decay could potentially receive a significant
contribution from long-distance effects, which may be comparable to that of
the short-distance effects.

%------------------------------------------------------------------------------
This work investigates the rescattering effects on $B_s\to K^{*0}\bar{K}^{*0}$
through triangle-rescattering diagrams, obtaining the contributions of
long-distance interactions to the branching ratio and longitudinal
polarization. It is pointed out that experimental observations of $b\to s$
processes can be well explained within the SM.

%------------------------------------------------------------------------------
\section{Formalism}
The $D_s^{(\ast)+}D_s^{(\ast)-}$ states from $B_s$ decays can rescatter to
$K^{*0}\bar{K}^{*0}$ through the $D^{+(\ast)}$ exchange in the triangle
diagrams, as depicted in Fig.~\ref{ku}. There are eight types of triangle
diagrams, $D_s^+D_s^-D$, $D_s^+D_s^-D^\ast$, $D_s^+D_s^{\ast-}D$,
$D_s^+D_s^{\ast-}D^\ast$, $D_s^{\ast+}D_s^-D$, $D_s^{\ast+}D_s^-D^\ast$,
$D_s^{\ast+}D_s^{\ast-}D$ and $D_s^{\ast+}D_s^{\ast-}D^\ast$. Each triangle
diagram have three vertices: one vertex involves the weak interaction~(marked
by blue), and the other two vertices involve the strong interaction (marked by
red). These two strong interaction vertices have equivalent coupling and can be
described with the same parametrization.

%------------------------------------------------------------------------------
Figure~\ref{ku}(a) shows the triangle diagrams at the quark level. The weak
interaction amplitudes for $B_s\to D_s^{(*)+}D_s^{(*)-}$ are dominated by the
color-allowed external $W$-emission and can be treated using QCD Factorization.
Following the approach described in
Refs.~\cite{Wu:2023fyh,Cheng:2003sm,Soni:2021fky}, we derive this amplitude to
be
\begin{eqnarray}\label{weakamp1}
{\cal M}(B_s\to D_s^{+}D_s^{-})
&=&B_1=-i\frac{G_f}{\sqrt{2}}V_{cb}V^{*}_{cs}a_1 f_{D_s}[m_{D_s}^2F_{-}^{B_sD_s}(p_1^2)+(m_{B_s}^2-m_{D_s}^2)F_{+}^{B_sD_s}(p_1^2)]\,,\nonumber\\
{\cal M}(B_s\to D_s^{*+}D_s^{-})&=&\epsilon_\mu(p_1)B_2^\mu
=\frac{2G_f}{\sqrt{2}}V_{cb}V^{*}_{cs}a_1 f_{D^*_s}m_{D^*_s}p_{2}.\epsilon(p_1)F_{+}^{B_sD_s}(p_1^2)\,,\nonumber\\
{\cal M}(B_s\to D_s^{+}D_s^{*-})&=&\epsilon_\mu(p_2)B_3^\mu\nonumber\\
&=&-i\frac{G_f}{\sqrt{2}}V_{cb}V^{*}_{cs}a_1 f_{D_s}p_{1}.\epsilon(p_2)[(m_{B_s}-m_{D^*_s})A_{+}^{B_sD_s^*}(p_1^2)+\frac{m_{D*_s}^2}{m_{B_s}+m_{D^*_s}}A_{-}^{B_sD_s^*}(p_1^2)]\,,\nonumber\\
{\cal M}(B_s\to D_s^{*+}D_s^{*-})&=&\epsilon_\mu(p_1)\epsilon_\nu(p_2)B_4^{\mu\nu}\nonumber\\
&=& -\frac{G_f}{\sqrt{2}}V_{cb}V^{*}_{cs}a_1 f_{D^*_s}\frac{m_{D^*_s}}{m_{B_s}+m_{D^*_s}}\epsilon^{\mu}(p_1)\epsilon^{\nu}(p_2)[-2i\epsilon_{\mu\nu\alpha\beta}p_2^{\alpha}p_1^{\beta}V_{0}^{B_sD_s^*}(p_1^2)\nonumber\\
   &+&(m_{B_s}-m_{D^*_s})^2g_{\mu\nu}A_{0}^{B_sD_s^*}(p_1^2)-(p_1+2p_2)_\mu(p_1+2p_2)_\nu A_{+}^{B_sD_s^*}(p_1^2)]\,,
\end{eqnarray}
where $p_{1},p_{2},G_F$, $V_{ij}$, $f_{D^{(*)}_s}$ and
$(F_{\pm}, A_{\pm, 0}, V_0)$ are the momenta of $D_s^{(\ast)+},D_s^{(\ast)-}$,
Fermi constant, Cabibbo-Kobayashi-Maskawa~(CKM) matrix elements, decay
constants of $D^{(*)}_s$, and $B_s\to D_{s}^{(*)}$ transition form factors,
respectively, while $a_1$ is the parameter related to the Wilson coefficients
from the factorization of $B_s\to D_s^{(*)+}D_s^{(*)-}$.

%------------------------------------------------------------------------------
The strong interaction amplitude for
$D_s^{(*)}\to D^{(*)}K^{*0}$~($\bar{D}_s^{(*)}\to D^{(*)}\bar{K}^{*0}$), using
the chiral and heavy quark symmetries~\cite{Cheng:2004ru,Casalbuoni:1996pg,Wu:2023fyh},
are given as:
\begin{eqnarray}
{\cal M}(D^{+}_s\to D^{+}K^{0*}) &=& -ig_{DDV}(p_1+q)\cdot\epsilon(p_3),
\nonumber\\
{\cal M}(D^{+}_s\to D^{+*}K^{0*}) &=& 2ig_{D^*DV} \epsilon_{\mu\nu\alpha\beta} p^\mu_3 \epsilon^\nu(p_3)
\cdot(p_1+q)^\alpha \epsilon^\beta(q) ,
\nonumber\\
{\cal M}(D^{+*}_s\to D^{+}K^{0*})&=&-2ig_{D^*DV}\epsilon_{\mu\nu\alpha\beta}p^{\mu}_3\epsilon^\nu(p_3)
(p_1+q)^{\alpha}\epsilon^\beta(p_1),
\nonumber\\
{\cal M}(D^{+*}_s\to D^{+*}K^{0*})& =& i \{ g_{D^*\!D^*\!V} (p_1\!+\!q) \!\cdot\! \epsilon(p_3) \epsilon(q) \!\cdot\! \epsilon(p_1)\nonumber\\
&&-4f_{D^*D^*V}[p_3\cdot\epsilon(q)\epsilon(p_3)\cdot\epsilon(p_1)
-p_3\cdot\epsilon(p_1)\epsilon(p_3)\cdot\epsilon(q)]\},
\end{eqnarray}
%------------------------------------------------------------------------------
The four coupling constants are expressed as
$g_{DDV} = g_{D^*D^*V}=\beta g_V / \sqrt2$ and
$f_{D^*DV}=f_{D^*D^*V}/m_{D^*}=\lambda_V g_V/\sqrt2$. The parameters $g_V$,
$\beta$, and $\lambda$ thus enter into the effective chiral Lagrangian
describing the interactions of heavy mesons with low-momentum light vector
mesons. Next, we adopt the optical theorem and Cutkosky cutting
rule~\cite{Cheng:2004ru,Yu:2020vlt,Han:2021azw}\footnote{The amplitude of
long-distance rescattering can also be calculated with the 'Hooft-Veltman
technique~\cite{tHooft:1978jhc, Wu:2023fyh, Hsiao:2019ait,Yu:2021euw,Yu:2022lwl}}
to compute the triangle diagrams in Fig.~\ref{ku}. The amplitude of
long-distance rescattering contributions for the decay
$B_s\to K^{*0}\bar{K}^{*0}$ is obtained as:
\begin{eqnarray}
  {\cal M}_{\rm LD}(B_s\to K^{*0}\bar{K}^{*0}) &=& \sum{\cal A}bs(D_s^{(*)+}D_s^{(*)-};D^{(*)}) ,
  \nonumber\\
      {\cal A}bs(D_s^{(*)+}D_s^{(*)-};D^{(*)}) &=& \frac{1}{2}\int \frac{d^{3} \vec{p}_1}{(2\pi)^32E_1}\frac{d^{3} \vec{p}_2}{(2\pi)^32E_2}(2\pi)^{4}
      \delta^{4}(P_{B_s}-p_1-p_2)\nonumber\\
      &&\times\sum_\lambda {\cal M}(B_s\to D_s^{(*)+}D_s^{(*)-}){\cal M}(D^{(*)+}_s\to D^{(*)+}K^{0*})\nonumber\\
      && \times{\cal M}(D^{(*)-}_s\to D^{(*)-}\bar{K}^{0*})\frac{F^2(q^2,m_q)}{q^2-m^2_q}\,.
      \label{w2}
\end{eqnarray}
%------------------------------------------------------------------------------
We have assumed that the exchanged $D^{(*)}$ is off-shell while $D_s^{(*)+}$
and $D_s^{(*)-}$ are on-shell, with the momentum angle between $D_s^{(*)+}$ and
the final state particle $K^{*0}$ given by $\cos\theta$. The form factor
$F(q^2,\Lambda) \equiv \frac{\Lambda^{2} - m_{q}^{2}}{\Lambda^{2} - q^{2}}$
takes care of the off-shell effect of the exchanged particle $D^{(*)}$.
Note that a cutoff $\Lambda$ must be introduced to the vertex to render the
whole calculation meaningful. The decay amplitudes
${\cal A}bs(D_s^{(*)+}D_s^{(*)-};D^{(*)})$ correspond to the eight triangle
diagrams related to the intermediate state $D_s^{(*)+}$, $D_s^{(*)-}$ and
$D^{(*)}$, which are written as:
\begin{eqnarray}\label{abs1}
  {\cal A}bs(D_s^{+}D_s^{-};D)
  &=& \frac{1}{2}\int \frac{d^{3} \vec{p}_1}{(2\pi)^32E_1}\frac{d^{3} \vec{p}_2}{(2\pi)^32E_2}(2\pi)^{4}
  \delta^{4}(P_{B_s}-p_1-p_2)B_1\nonumber\\
  &&(-i)g_{DDV}(p_1+q)\cdot\epsilon_3(-i)g_{DDV}(p_2-q)\cdot\epsilon_4\frac{iF^2(q^2,\Lambda)}{q^2-m^2_q}\nonumber\\
  &=& -i\int^{+1}_{-1}d\cos\theta\frac{|\vec{p_1}|B_1g^2_{DDV}}{16\pi m_{B_s}}
  (p_1+q)\cdot\epsilon_3(p_2-q)\cdot\epsilon_4\frac{F^2(q^2,\Lambda)}{q^2-m^2_{B_s}}\,,
\end{eqnarray}
\begin{eqnarray}\label{abs2}
  {\cal A}bs(D_s^{+}D_s^{-};D^\ast)
  &=& \frac{1}{2}\int \frac{d^{3} \vec{p}_1}{(2\pi)^32E_1}\frac{d^{3} \vec{p}_2}{(2\pi)^32E_2}(2\pi)^{4}
  \delta^{4}(P_{B_s}-p_1-p_2)B_1\nonumber\\
  &&\times(-2i)f_{D^\ast DV}\epsilon_{\mu\nu\alpha\beta}(ip_3^\mu)\epsilon^\nu_3(i)(p_1+q)^\alpha (-2i)f_{D^\ast DV}\nonumber\\
  &&\times\epsilon_{\mu^\prime\nu^\prime\alpha^\prime\beta^\prime}ip_4^{\mu^\prime}\epsilon^{\nu^\prime}_4(-i)(q-p_2)^{\alpha^\prime}(-g^{\beta\beta^\prime}+\frac{q^{\beta}q^{\beta^\prime}}
        {m_{D^\ast}^2})\frac{iF^2(q^2,\Lambda)}{q^2-m^2_q}\nonumber\\
        &=& i\int^{+1}_{-1}d\cos\theta\frac{|\vec{p_1}|B_1f^2_{D^\ast DV}}{\pi m_{B_s}}\epsilon_{\mu\nu\alpha\beta}p_3^\mu\epsilon^\nu_3p_1^\alpha\epsilon_{\mu^\prime\nu^\prime\alpha^\prime\beta^\prime}p_4^{\mu^\prime}\epsilon^{\nu^\prime}_4p_2^{\alpha^\prime}g^{\beta\beta^\prime}
        \frac{F^2(q^2,\Lambda)}{q^2-m^2_{D^\ast}}\,,
\end{eqnarray}
\begin{eqnarray}\label{abs3}
  {\cal A}bs(D_s^{\ast+}D_s^{-};D)
  &=& \frac{1}{2}\int \frac{d^{3} \vec{p}_1}{(2\pi)^32E_1}\frac{d^{3} \vec{p}_2}{(2\pi)^32E_2}(2\pi)^{4}
  \delta^{4}(P_{B_s}-p_1-p_2)B_{2\rho}\nonumber\\
  &&\times(-2i)f_{D^\ast DV}\epsilon_{\mu\nu\alpha\beta}(ip_3^\mu)\epsilon^\nu_3 (-i)(p_1+q)^\alpha (-i)g_{DDV}\nonumber\\
  &&\times (q-p_2)\cdot\epsilon_4(-g^{\beta\rho}+\frac{p_1^{\beta}p_1^{\rho}}{m_{D_s^\ast}^2})\frac{iF^2(q^2,\Lambda)}{q^2-m^2_D}\nonumber\\
  &=& -i\int^{+1}_{-1}d\cos\theta\frac{|\vec{p_1}|f_{D^\ast DV}g_{DDV}}{2\pi m_{B_s}}\epsilon_{\mu\nu\alpha\beta}p_3^\mu\epsilon^\nu_3p_1^\alpha p_2\cdot\epsilon_4B_{2}^{\beta}
  \frac{F^2(q^2,\Lambda)}{q^2-m^2_D}\,,
\end{eqnarray}
\begin{eqnarray}\label{abs4}
  {\cal A}bs(D_s^{\ast+}D_s^{-};D^\ast)
  &=& \frac{1}{2}\int \frac{d^{3} \vec{p}_1}{(2\pi)^32E_1}\frac{d^{3} \vec{p}_2}{(2\pi)^32E_2}(2\pi)^{4}
  \delta^{4}(P_{B_s}-p_1-p_2)B_{2\rho}\nonumber\\
  &&\times\{ig_{D^\ast D^\ast V}(p_1+q)\cdot\epsilon_3g^{\mu\nu}-4if_{D^\ast D^\ast V}[p_3^\mu \epsilon^\nu_3-p_3^\nu \epsilon^\mu_3]\} (-2i)f_{D^\ast D V}\nonumber\\
  &&\times\epsilon_{\mu^\prime\nu^\prime\alpha^\prime\beta^\prime}p_4^{\mu^\prime}\epsilon^{\nu^\prime}_4(q-p_2)^{\alpha^\prime}(-g^{\rho\nu}+\frac{p_1^{\rho}p_1^{\nu}}{m_{D_s^\ast}^2})
  (-g^{\mu\beta^\prime}+\frac{q^{\mu}q^{\beta^\prime}}{m_{D^\ast}^2})\frac{iF^2(q^2,\Lambda)}{q^2-m^2_{D^\ast}}\nonumber\\
  &=& i\int^{+1}_{-1}d\cos\theta\frac{|\vec{p_1}|f_{D^\ast DV}}{2\pi m_{B_s}}\{g_{D^\ast D^\ast V}p_1\cdot\epsilon_3g^{\mu\nu}-2f_{D^\ast D^\ast V}[p_3^\mu \epsilon^\nu_3-p_3^\nu \epsilon^\mu_3]\}\nonumber\\
  &&\times\epsilon_{\mu^\prime\nu^\prime\alpha^\prime\mu}p_4^{\mu^\prime}\epsilon^{\nu^\prime}_4p_2^{\alpha^\prime}(-g^{\rho\nu}+\frac{p_1^{\rho}p_1^{\nu}}{m_{D_s^\ast}^2})B_{2\rho}
  \frac{F^2(q^2,\Lambda)}{q^2-m^2_{D^\ast}}\,,
\end{eqnarray}
\begin{eqnarray}\label{abs5}
  {\cal A}bs(D_s^{+}D_s^{\ast-};D)
  &=& \frac{1}{2}\int \frac{d^{3} \vec{p}_1}{(2\pi)^32E_1}\frac{d^{3} \vec{p}_2}{(2\pi)^32E_2}(2\pi)^{4}
  \delta^{4}(P_{B_s}-p_1-p_2)B_{3\rho}\nonumber\\
  &&\times(-i)g_{DDV}(p_1+q)\cdot\epsilon_3 (-2i)f_{D^\ast D V}\nonumber\\
  &&\times\epsilon_{\mu\nu\alpha\beta}ip_4^{\mu}\epsilon^{\nu}_4i(q-p_2)^{\alpha}(-g^{\rho\beta}+\frac{p_2^{\rho}p_2^{\beta}}{m_{D_s^\ast}^2})
  \frac{iF^2(q^2,\Lambda)}{q^2-m^2_D}\nonumber\\
  &=& i\int^{+1}_{-1}d\cos\theta\frac{|\vec{p_1}|g_{DDV}f_{D^\ast DV}}{2\pi m_{B_s}}p_1\cdot\epsilon_3\epsilon_{\mu\nu\alpha\beta}p_4^{\mu}\epsilon^{\nu}_4p_2^{\alpha}B^\beta_{3}
  \frac{F^2(q^2,\Lambda)}{q^2-m^2_{D}}\,,
\end{eqnarray}
\begin{eqnarray}\label{abs6}
  {\cal A}bs(D_s^{+}D_s^{\ast-};D^\ast)
  &=& \frac{1}{2}\int \frac{d^{3} \vec{p}_1}{(2\pi)^32E_1}\frac{d^{3} \vec{p}_2}{(2\pi)^32E_2}(2\pi)^{4}
  \delta^{4}(P_{B_s}-p_1-p_2)B_{3\rho}\nonumber\\
  &&\times 2if_{D^\ast D V}\epsilon_{\mu\nu\alpha\beta}p_3^{\mu}\epsilon^{\nu}_3(q+p_1)^{\alpha}(-g^{\beta\sigma}+\frac{q^{\beta}q^{\sigma}}{m_{D^\ast}^2})(-g^{\rho\lambda}+\frac{p_2^{\rho}p_2^{\lambda}}{m_{D_s^\ast}^2})\nonumber\\
  &&\times
  \{ig_{D^\ast D^\ast V}(q-p_2)\cdot\epsilon_4g^{\lambda\sigma}-4if_{D^\ast D^\ast V}[p_4^\lambda \epsilon^\sigma_4-p_4^\sigma \epsilon^\lambda_4]\}\frac{iF^2(q^2,\Lambda)}{q^2-m^2_{D^\ast}}\nonumber\\
  &=&- i\int^{+1}_{-1}d\cos\theta\frac{|\vec{p_1}|f_{D^\ast DV}}{2\pi m_{B_s}}\{g_{D^\ast D^\ast V}p_2\cdot\epsilon_4g^{\lambda\sigma}+2f_{D^\ast D^\ast V}[p_4^\lambda \epsilon^\sigma_4-p_4^\sigma \epsilon^\lambda_4]\}\nonumber\\
  &&\times\epsilon_{\mu\nu\alpha\sigma}p_3^{\mu}\epsilon^{\nu}_3p_1^{\alpha}(-g^{\rho\lambda}+\frac{p_2^{\rho}p_2^{\lambda}}{m_{D_s^\ast}^2})B_{3\rho}
  \frac{F^2(q^2,\Lambda)}{q^2-m^2_{D^\ast}}\,,
\end{eqnarray}
\begin{eqnarray}\label{abs7}
  {\cal A}bs(D_s^{+}D_s^{\ast-};D^\ast)
  &=& \frac{1}{2}\int \frac{d^{3} \vec{p}_1}{(2\pi)^32E_1}\frac{d^{3} \vec{p}_2}{(2\pi)^32E_2}(2\pi)^{4}
  \delta^{4}(P_{B_s}-p_1-p_2)B_{4\rho\sigma}\nonumber\\
  &&\times 2if_{D^\ast D V}\epsilon_{\mu\nu\alpha\beta}p_3^{\mu}\epsilon^{\nu}_3(q+p_1)^{\alpha} 2if_{D^\ast D V}\epsilon_{\mu^\prime\nu^\prime\alpha^\prime\beta^\prime}p_4^{\mu^\prime}\epsilon^{\nu^\prime}_4(q-p_2)^{\alpha^\prime}\nonumber\\
  &&\times
  (-g^{\rho\beta}+\frac{p_1^{\rho}p_1^{\beta}}{m_{D_s^\ast}^2})(-g^{\sigma\beta^\prime}+\frac{p_2^{\sigma}p_2^{\beta^\prime}}{m_{D_s^\ast}^2})\frac{iF^2(q^2,\Lambda)}{q^2-m^2_{D}}\nonumber\\
  &=& i\int^{+1}_{-1}d\cos\theta\frac{|\vec{p_1}|f^2_{D^\ast DV}}{\pi m_{B_s}}\epsilon_{\mu\nu\alpha\beta}p_3^{\mu}\epsilon^{\nu}_3p_1^{\alpha} \epsilon_{\mu^\prime\nu^\prime\alpha^\prime\beta^\prime}p_4^{\mu^\prime}\epsilon^{\nu^\prime}_4p_2^{\alpha^\prime}B_{4}^{\beta\beta^\prime}
  \frac{F^2(q^2,\Lambda)}{q^2-m^2_{D^\ast}}\,,
\end{eqnarray}
\begin{eqnarray}\label{abs8}
  {\cal A}bs(D_s^{+}D_s^{\ast-};D^\ast)
  &=& \frac{1}{2}\int \frac{d^{3} \vec{p}_1}{(2\pi)^32E_1}\frac{d^{3} \vec{p}_2}{(2\pi)^32E_2}(2\pi)^{4}
  \delta^{4}(P_{B_s}-p_1-p_2)B_{4\rho\sigma}\nonumber\\
  &&\times \{ig_{D^\ast D^\ast V}(p_1+q)\cdot\epsilon_3g^{\mu\nu}-4if_{D^\ast D^\ast V}[p_3^\mu \epsilon^\nu_3-p_3^\nu \epsilon^\mu_3]\}\nonumber\\
  &&\times
  \{ig_{D^\ast D^\ast V}(q-p_2)\cdot\epsilon_4g^{\alpha\beta}-4if_{D^\ast D^\ast V}[p_4^\alpha \epsilon^\beta_4-p_4^\beta \epsilon^\alpha_4]\}\nonumber\\
  && (-g^{\rho\nu}+\frac{p_1^{\rho}p_1^{\nu}}{m_{D_s^\ast}^2})(-g^{\sigma\alpha}+\frac{p_2^{\sigma}p_2^{\alpha}}{m_{D_s^\ast}^2})(-g^{\mu\beta}+\frac{q^{\mu}q^{\beta}}{m_{D^\ast}^2})\frac{iF^2(q^2,\Lambda)}{q^2-m^2_{D^\ast}}\nonumber\\
  &=& i\int^{+1}_{-1}d\cos\theta\frac{|\vec{p_1}|}{4\pi m_{B_s}}\{g_{D^\ast D^\ast V}p_1\cdot\epsilon_3g^{\mu\nu}-2f_{D^\ast D^\ast V}[p_3^\mu \epsilon^\nu_3-p_3^\nu \epsilon^\mu_3]\}\nonumber\\
  &&\times
  \{g_{D^\ast D^\ast V}p_2\cdot\epsilon_4g^{\alpha\beta}+2f_{D^\ast D^\ast V}[p_4^\alpha \epsilon^\beta_4-p_4^\beta \epsilon^\alpha_4]\}\nonumber\\
  &&\times (-g^{\rho\nu}+\frac{p_1^{\rho}p_1^{\nu}}{m_{D_s^\ast}^2})(-g^{\sigma\alpha}+\frac{p_2^{\sigma}p_2^{\alpha}}{m_{D_s^\ast}^2})(-g^{\mu\beta}+\frac{q^{\mu}q^{\beta}}{m_{D^\ast}^2})B_{4\rho\sigma}\frac{F^2(q^2,\Lambda)}{q^2-m^2_{D^\ast}}\,.
\end{eqnarray}
%------------------------------------------------------------------------------
In order to obtain a general form of the amplitude of long-distance as
\begin{align}\label{eq:general}
  &{\cal M}_{\rm LD}(B_s\to K^{*0}\bar{K}^{*0}) = i\epsilon_{\mu}(K^{*0})\epsilon_{\nu}(\bar{K}^{*0})
  (g^{\mu\nu}S_1+p_{B_s}^{\mu}p_{B_s}^{\nu}S_2+i\epsilon^{\mu\nu\alpha\beta}p_{3\alpha}p_{B_s\beta}S_3)\,,
\end{align}
where $S_{1,2,3}$ will be functions of $\Lambda$. We first use the relationship,
$p_2=p_{B_s}-p_1,p_4=p_{B_s}-p_3$ and $q=p_{1}-p_3$, to write
Eqs.~(\ref{abs1})-(\ref{abs8}) in terms of $p_1$, $p_3$, $p_{B_s}$, $\epsilon_3$,
$\epsilon_4$ and $\Lambda$. The inner products $p_1\cdot p_3$, $p_{B_s}\cdot p_1$,
and $p_{B_s}\cdot p_3$ in Eqs.~(\ref{abs1})-(\ref{abs8}) can be expressed as
follows with the assumption that $D_s^{(*)\pm}$ are on-shell:
\begin{eqnarray}
  p_1\cdot p_3 &=& E_1E_3-|\vec{p_1}| |\vec{p_3}|\cos\theta,\,\,p_{B_s}\cdot p_1 = E_1m_{B_s},\,\,p_{B_s}\cdot p_3 = E_3m_{B_s}\,,
\end{eqnarray}
where
\begin{eqnarray}
|\vec{p_1}| &=& \frac{\sqrt{[m^2_{B_s}-(m_1+m_2)^2][m^2_{B_s}-(m_1-m_2)^2]}}{2m_{B_s}},\,\, E_1=|\vec{p_1}|^2+m_1^2,\,\,\nonumber\\
|\vec{p_3}| &=& \frac{\sqrt{[m^2_{B_s}-(m_3+m_4)^2][m^2_{B_s}-(m_3-m_4)^2]}}{2m_{B_s}},\,\, E_3=|\vec{p_3}|^2+m_3^2\,.
\end{eqnarray}
%------------------------------------------------------------------------------
Now, Eqs.~\ref{abs1}-\ref{abs8} are linear combinations of the following terms,
which we classify into three parts: the first part is
$\epsilon_3\cdot\epsilon_4$, $\epsilon_3\cdot p_{B_s}\epsilon_4\cdot p_{B_s}$
and
$\epsilon_{\mu\nu\alpha\beta}\epsilon^\mu_3\epsilon^\nu_4p^\alpha_{3}p^\beta_{B_s}$;
the second part is $\epsilon_3\cdot p_{1}\epsilon_4\cdot p_{B_s}$,
$\epsilon_3\cdot p_{B_s}\epsilon_4\cdot p_{1}$,
$\epsilon_{\mu\nu\alpha\beta}\epsilon^\mu_3\epsilon^\nu_4p^\alpha_{1}p^\beta_{B_s}$
and
$\epsilon_{\mu\nu\alpha\beta}\epsilon^\mu_3\epsilon^\nu_4p^\alpha_{3}p^\beta_{1}$;
the third part is $\epsilon_3\cdot p_{1}\epsilon_4\cdot p_{1}$. The second and
third parts can be further expressed as a linear combination of the first
part~\cite{Cheng:2003sm,Cheng:2005bg} with coefficients as functions of
$\cos\theta$. Here, these relations are used:
\begin{eqnarray}
  p_1^\mu &=& p_{B_s}^\mu A_1^{(1)}+(2p_{3}-p_{B_s})^\mu A_2^{(1)}\,,
\end{eqnarray}
and
\begin{eqnarray}
  p_1^\mu p_1^\nu &=& g^{\mu\nu} A_1^{(2)}+p_{B_s}^\mu p_{B_s}^\nu A_2^{(2)}+[p_{B_s}^\nu(2p_{3}-p_{B_s})^\mu +p_{B_s}^\mu(2p_{3}-p_{B_s})^\nu]A_3^{(2)}\nonumber\\
  &&+(2p_{3}-p_{B_s})^\mu(2p_{3}-p_{B_s})^\nu A_4^{(2)}\,,
\end{eqnarray}
where
\begin{eqnarray}
  A_1^{(1)} &=& \frac{p_{B_s}\cdot p_{1}}{m^2_{B_s}},\,\, A_2^{(1)} =\frac{2p_{3}\cdot p_{1}-p_{B_s}\cdot p_{1}}{4m^2_{K^\ast}-m^2_{B_s}}\nonumber\\
  \begin{bmatrix}
  A_1^{(2)} \\
  A_2^{(2)} \\
  A_3^{(2)} \\
  A_4^{(2)}
\end{bmatrix} &=&\begin{bmatrix}
  4 & m^2_{B_s} & 0 & q^2 \\
  m^2_{B_s} & m^4_{B_s} & 0 & 0 \\
  0 & 0 & 2m^2_{B_s}q^2 & 0 \\
  q^2 & 0 & 0 & (q^2)^2
\end{bmatrix}^{-1}\begin{bmatrix}
  p^2_1 \\
  (p_{B_s}\cdot p_1)^2 \\
  2p_{B_s}\cdot p_1q\cdot p_1 \\
 [q\cdot p_1]^2
\end{bmatrix}
\end{eqnarray}
with $q=2p_3-p_{B_s}$ and $q^2=4m^2_{K^\ast}-m^2_{B_s}$. Eventually, the general
form of the amplitude of long-distance, Eq.~\ref{eq:general}, can be derived and
$S_{1,2,3}$ can be determined by integrating over $\cos\theta$ in
Eq.~(\ref{abs1}-\ref{abs8}), which are the function of $\Lambda$.
Consequently,
the decay width $\Gamma_{\rm LD}$, longitudinal polarization $f_{\rm L, LD}$,
and perpendicular polarization $f_{\rm \perp, LD}$ are derived to be
\begin{align}
&\Gamma_{\rm LD}(B_s\to K^{*0}\bar{K}^{*0}) =\frac{ p_c}{8\pi m^2_{B_s}}
 \bigg[\Big|S_1+\frac{m^2_{B_s}}{2}S_2\Big|^2+|S_1|^2+\Big|\frac{m^2_{B_s}-2m^2_{K^{*0}}}{2}S_3\Big|^2\bigg]\,,
\label{Eq:GammaLD}
\\
& f_{\rm L, LD}^{\rm }(B_s\to K^{*0}\bar{K}^{*0}) =\frac{ p_c}{8\pi m^2_{B_s}}\Big|S_1 + \frac{m^2_{B_s}}{2} S_2\Big|^2
 \bigg[\Gamma_{\rm LD}(B_s\to K^{*0}\bar{K}^{*0})\bigg]^{-1}\,,
 \label{Eq:fLLD}
 \\
 & f_{\rm \perp, LD}^{\rm }(B_s\to K^{*0}\bar{K}^{*0}) =
 \frac{ p_c}{16\pi m^2_{B_s}}\Big|S_1-\frac{m^2_{B_s}-2m^2_{K^{*0}}}{2}S_3\Big|^2
 \bigg[\Gamma_{\rm LD}(B_s\to K^{*0}\bar{K}^{*0})\bigg]^{-1}\,.
\label{Eq:fPLD}
\end{align}
%------------------------------------------------------------------------------

\section{Numerical results}
To estimate the contributions from rescattering amplitudes, we need to specify
various parameters entering into the vertices of Feynman diagrams. The decay
constants is extracted by BESIII to be $f_{D^*_s}\simeq f_{D_s}=0.249$~GeV in
the $D_s^+\to\mu^+\nu_\mu$ decay~\cite{Ke:2023qzc}.
The parameters relevant for the $D_s^{(*)}\to D^{(*)}K^{*0}$ strong coupling
are $(g_V, \beta, \lambda_V)=(5.8,\,0.9,\,0.56~\text{GeV})$~\cite{Cheng:2004ru}.
The parameter, $a_1\simeq1.0$, related to short-distance factorizable
amplitudes is commonly expected to be close to 1~\cite{Chua:2019yqh}. The form
factors are calculated with the covariant confined quark model as
$F_i(q^2)=\frac{F_i(0)}{1-a\frac{q^2}{m_{B_s}^2}+b(\frac{q^2}{m_{B_s}^2})^2}$
with $F_i=(F_+,F_-,A_+,A_-,A_0,V_0)$. The resultant values~\cite{Soni:2021fky}
are listed in Table~\ref{tab1}. For the CKM matrix elements, we use the
world-average values from PDG~\cite{ParticleDataGroup:2022pth},
$(V_{cb},V_{cs})=(A\lambda^2,1-\lambda^2/2)$, with $A= 0.790\pm 0.017$ and
$\lambda=0.22650\pm 0.00048$. The cutoff parameter should be not far from the
physical mass of the exchanged $D^{(*)}$ and is given as
$\Lambda=m_{D^{(*)}}+\eta\Lambda_{\rm QCD}$, with
$\Lambda_{\rm QCD} = 0.22$~GeV~\cite{Cheng:2004ru}. The energy scale $\eta$ is
introduced to adjust for the theoretical assumptions and the associated errors.

%------------------------------------------------------------------------------
\begin{table}
\footnotesize
\caption{The values of the parameters $F_i(0)$, $a$, and $b$ in the form
  factors for  $B_s\to D^{(*)}_{s}$}
\label{tab1}
\begin{tabular}{lcccccc}
  \hline
  & $F_+$ &$ F_-$ &$ A_+$ & $A_-$ & $A_0$ &$V_0$ \\ \hline
  $F_i(0)$ &0.770 & -0.355 & 0.630 & -0.756 & 1.564  & 0.743  \\
  $a$ &  0.837 &  0.855 &  0.972  &  1.001  & 0.442  & 1.010  \\
  $b$ &  0.077 &  0.083 & -0.092 & 0.116 & -0.178 & 0.118 \\
  \hline
\end{tabular}
\end{table}

\begin{figure}[t!]
\centering
\includegraphics[width=2.0in]{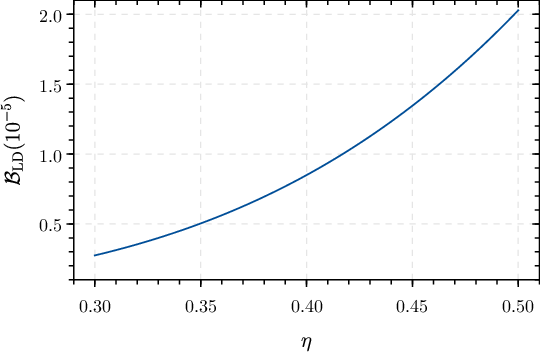}
\includegraphics[width=2.0in]{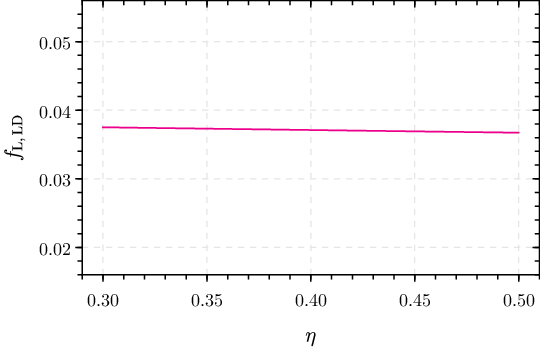}
\includegraphics[width=2.0in]{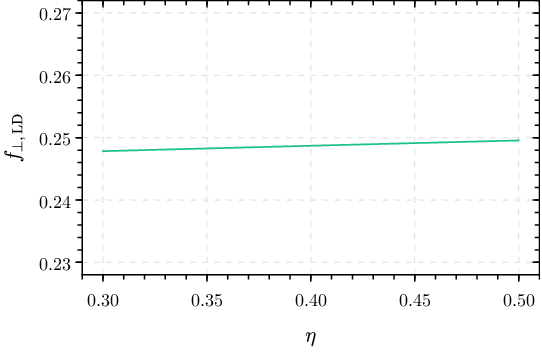}
\caption{The branching ratio, longitudinal polarization, and perpendicular polarization
  by long-distance contributions related to $\eta$.}\label{triangle}
\end{figure}

%------------------------------------------------------------------------------
We then display the dependence of the branching ratio, longitudinal
polarization, and perpendicular polarization by long-distance contributions on
$\eta$ as shown in Fig.~\ref{triangle}. It is intriguing to observe that the
branching ratio contributed by long-distance interactions is of the same order
of magnitude as that of short-distance interactions, and the impact on
longitudinal and perpendicular polarizations from long-distance interactions can not be ignored.
While the branching ratio of long-distance interactions is quite sensitive to
$\eta$, the dependence of longitudinal and perpendicular
polarization rates on $\eta$ is not significant and the rate remains
consistently small. Note that the longitudinal and perpendicular
amplitudes positively correlate with the total amplitude, resulting in the
cancellation of the dependence on $\eta$, as defined in
Eqs.~(\ref{Eq:fLLD}) and (\ref{Eq:fPLD}). 

To effectively compare with experimental data, the branching ratio and
longitudinal (perpendicular) polarization with contributions from
both long-distance and short-distance are expressed as
\begin{align}
&{\cal B}= {\cal B}_{\rm LD}+{\cal B}_{\rm SD}\,,\nonumber\\
&f_{\rm L(\perp)}=\bigg[{\cal B}_{\rm LD} f_{\rm L(\perp), LD}+{\cal B}_{\rm SD}f_{\rm L(\perp),SD}\bigg]{\cal B}^{-1}\,.
\end{align}
The interference between the long- and short-distance interactions is
considered negligible, as the long-distance contribution is essentially
``orthogonal'' to the short-distance one. The short-distance interaction
amplitude for $B \to \phi K^*$ is purely penguin-like with a weak phase arising
from $V_{tb} V_{ts}^*$. Along with the unitarity relation of the CKM matrix,
$V_{tb} V_{ts}^* = -V_{cb} V_{cs}^* - V_{ub} V_{us}^*$, and
$|V_{cb} V_{cs}^*| \gg |V_{ub} V_{us}^*|$, one can conclude that the weak
phases of the short- and long-distance are both dominated by $V_{cb} V_{cs}^*$.
Given that $A = A_{\rm SD} + iA_{\rm LD}$, when the weak phases of the short-
and long-distance interaction are aligned by $V_{cb} V_{cs}^*$, the
long-distance contribution becomes effectively orthogonal to the short-distance
one~\cite{Cheng:2005bg}. Furthermore, the largest effective Wilson coefficient
$a_4$, as demonstrated in Ref.~\cite{Chen:1998dta}, has a
significantly smaller imaginary component compared to its real counterpart,
leading to $a_4$ being approximately a real number.

We utilize the branching ratio
${\cal B}_{\rm SD} = (6.68^{+2.9}_{-2.2})\times10^{-6}$, the longitudinal
polarization $f_{\rm L,SD} = 0.464^{+0.127}_{-0.129}$, and the
perpendicular polarization $f_{\rm \perp,SD} = 0.235^{+0.058}_{-0.059}$
from short-distance calculation in Ref.~\cite{Yan:2018fif} and set
$\eta=\,0.35$.
Subsequently, the the branching ratio, longitudinal polarization, and
perpendicular polarization are determined to be
\begin{eqnarray}
{\cal B}&=&(1.17^{+0.29}_{-0.22})\times10^{-5},\,\,f_{\rm L}=0.28^{+0.06}_{-0.06}\,,f_{\rm \perp}=0.24^{+0.06}_{-0.05}\,,
\end{eqnarray}
which is consistent with the latest averaged experimental
results~\cite{ParticleDataGroup:2022pth}
${\cal B}(B_s\to K^{*0}\bar{K}^{*0})=(1.11\pm0.27)\times10^{-5}$,
$f_{\rm L}(B_s\to K^{*0}\bar{K}^{*0})=0.240\pm0.031\pm0.025$, and
$f_{\rm \perp}(B_s\to K^{*0}\bar{K}^{*0})=0.38\pm0.11\pm0.04$.
Moreover, the parameter $\eta$ is unconstrained, leading to significant
uncertainties in calculations within the SM. These uncertainties present
significant obstacles to utilizing this decay channel for the purpose of
searching for NP.

%------------------------------------------------------------------------------
\section{Summary}
We have studied the effects of final-state interactions on the branching ratio
and longitudinal polarization of $B_s\to K^{*0}\bar{K}^{*0}$. Ignoring the
long-distance interaction may lead to underestimate the effects of terms
involving the CKM matrix element $V_{cb}^*V_{cs}$ in the decay amplitude.
Our calculation suggests that the search for NP in $B \to VV$ processes can
not ignore the significant influence of long-distance interactions.
%process and the large theoretical uncertainty implies that $B \to VV$ may not
%be the golden mode for searching for NP.
These long-distance effects may also contribute
significantly to other decay processes, such as $B_s\to K^{0}\bar{K}^{*0}$,
$B_s\to K^{*0}\bar{K}^{0}$, $B_s\to K^{0}\bar{K}^{0}$, $B_s\to K^{+}K^{-}$,
$B_d\to \phi K^{(*)0}$, and more.
In our forthcoming article, we will delve into these decay processes in further
detail.
%------------------------------------------------------------------------------

%{\it Acknowledgments.--}
YY was supported in part by National Natural Science Foundation of China~(NSFC) under Contracts No.~11905023, No.~12047564 and No.~12147102, and the Science and Technology Research Program of Chongqing Municipal Education Commission (STRPCMEC) under Contracts No.~KJQN202200605 and No.~KJQN202200621; HBF was supported in part by NSFC under Contracts No.~12265010; HZ and BCK were supported in part by the Excellent Youth Foundation of Henan Scientific Committee under Contract No.~242300421044, NSFC under Contracts No.~11875054 and No.~12192263, and Joint Large-Scale Scientific Facility Fund of the NSFC and the Chinese Academy of Sciences under Contract No.~U2032104.

\end{document}